# NUMERISATION D'UN SIECLE DE PAYSAGE FERROVIAIRE FRANÇAIS :

## recul du rail, conséquences territoriales et coût environnemental

**Robert JEANSOULIN**, LIGM UMR8049, Université Gustave Eiffel, Marne-la-Vallée

Contact : robert.jeansoulin@univ-eiffel.fr

# NUMERISATION D'UN SIECLE DE PAYSAGE FERROVIAIRE FRANÇAIS :

## recul du rail, conséquences territoriales et coût environnemental

Robert JEANSOULIN, LIGM UMR8049, Université Gustave Eiffel, Marne-la-Vallée

Contact : robert.jeansoulin@univ-eiffel.fr

**Résumé**

La reconstitution de données géographiques sur une longue période permet de visualiser l'évolution du paysage ferroviaire français sur un siècle, au gré des grands événements de courte durée (guerre) ou de plus longue durée (délocalisation de secteurs industriels, métropolisation). Ce travail est le fruit d'opérations de fusion d'informations géographiques provenant de sources publiques (SNCF, IGN) et de la collecte assistée par ordinateur de sources « volontaires » ouvertes sur l'Internet (Wikipédia et sites amateurs). Les données numériques codées sont les gares (y compris simples arrêts) et nœuds d'embranchement associés à leurs lignes respectives, correctement géocodés et ordonnés afin de reconstituer le graphe cohérent du réseau. Les dates de validité des lignes permettent d'en déduire l'état à une date donnée. L'état actuel de la reconstitution, environ 90% du total, permet d'envisager certaines évaluations démographiques (communes servies par le réseau ferré entre 1925 et aujourd'hui) et environnementales (émissions de $CO^2$ simulées par trajet).

**Mots clefs**

SIG historique, fusion de données géographiques, information géographique volontaire, patrimoine ferroviaire, coût environnemental du transport

## Introduction

Le rapport (EU, 2016) ne note pas d'inflexion majeure, dans les politiques des états membres, à la hauteur des recommandations du Livre Blanc Européen 2010 sur les transports. Le mouvement des gilets jaunes a pointé l'absence d'alternative à l'automobile pour la « France des ronds-points ». La santé du tissu économique va de pair avec le maillage du réseau ferré : les fermetures de lignes SCNF suivent les fermetures d'usines comme une double peine. Sur les grands enjeux, la « politique » résulte d'une série de décisions dont les conséquences s'échelonnent sur plusieurs générations.

La documentation d'un siècle de paysage ferroviaire français, exposée dans cet article, a l'ambition d'offrir un outil d'analyse rétrospectif pour évaluer ou simuler des décisions alternatives. La collecte des données ferroviaires sous forme numérique est réalisée par les services publics, pour ce qui concerne les infrastructures officielles, actuellement en service ou encore sous responsabilité de la SNCF. De nombreuses lignes secondaires échappent à ce recensement, mais nombre de travaux d'amateurs permettent d'en reconstituer le paysage : géographie volontaire, ou participative comme Wikipedia. Selon (Auphan, 2011) la longueur du réseau ferré français a atteint 70000 km au début des années *1920*, sans mention du nombre de gares ou de communes desservies par ce réseau : les deux tiers ont disparu aujourd'hui, ce qui montre l'étendue du travail de reconstitution.

Les outils actuels de collecte automatique ou semi-automatique des données sur la toile permettent de glaner l'information recherchée. Cet article expose comment géo-localiser l'ensemble des gares ayant existé de 1925 à nos jours, et comment le graphe du réseau de ces gares permet d'effectuer des calculs de chemins, des calculs sur la mobilité ferroviaire –virtuelle-, et des calculs sur l'efficacité énergétique du transport des personnes et des marchandises.

La section 1 expose la méthodologie utilisée dans (Jeansoulin, 2021) pour reconstituer ce réseau sous forme d'un fichier des gares géocodées et connectées : fusion de données issues de plusieurs sources, extraction de données collectées sur Wikipedia ou sites de données géographiques volontaires (crowdsourcing). La section 2 présente l'état actuel de cette collecte et fournit quelques statistiques sur la qualité des données consolidées. La section 3 présente les outils pour effectuer des évaluations ou simulations qu'il est possible de développer dans plusieurs domaines des sciences sociales (mobilités alternatives), ou de l'environnement (calculs alternatifs d'émission $CO^2$).

## 1 Les outils numériques de la reconstitution du réseau

### 1.1 La cible : un graphe de gares connectées et géo-localisées

L'objectif est de situer sur le territoire les gares, haltes ou simples arrêts facultatifs ayant existé en France depuis l'apogée du réseau ferré à la sortie de la guerre 14-18. La géo-localisation donnera accès aux communes concernées et la séquences des gares d'une ligne permettra le calcul des distances et des chemins à parcourir sur le graphe du réseau. Le modèle 'INSPIRE' a été développé au niveau européen (INSPIRE,2012), afin de représenter un tel graphe, illustré Fig.1.

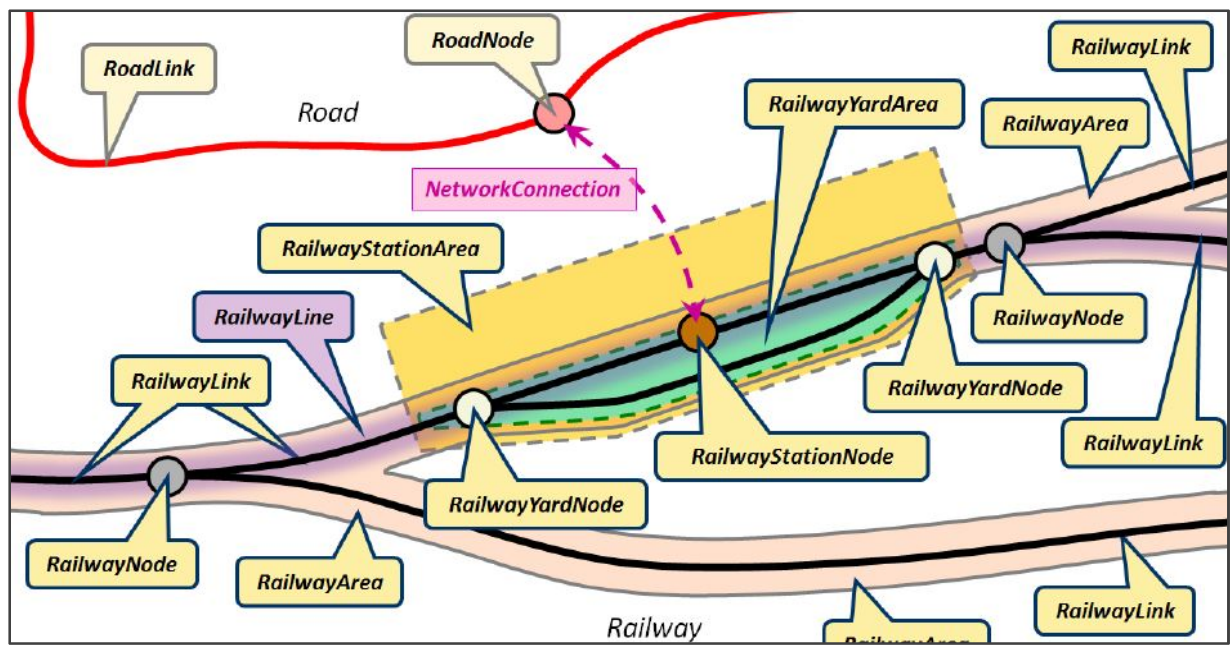

*Figure 1 : modèle standard européen « INSPIRE » : noter la différence entre RailwayNode, RailwayStationNode et RailwayYardNode, ainsi que la notion de NetworkConnection.*

Pour atteindre l'objectif il faut donc

- (a) identifier le maximum de gares et les géo-localiser,
- (b) connaitre les lignes qui relient ces gares dans un ordre déterminé,
- (c) associer les gares à leurs communes par géocodage.

### 1.2     Données ouvertes (publiques, volontaires, participatives)

#### *1.2.1   Données publiques*

Trois sources de données publiques permettent d'atteindre les objectifs (a-c) : provenant de (SNCF, 2020) et (IGN, 2018), voir Tab.1. Pour la SNCF, il existe des copies de versions antérieures (conservées sans curation) qui seront utilisées en partie « Fusion » de la section 2.

| origine | versions | #items | géométrie |
|---|---|---|---|
| SNCF_Gares | 2020 (+ 2017, 2014) | 4148  (6812, 6442) | points |
| SNCF_Lignes | 2019 | 801 segments | lignes |
| IGN_Communes | 2015 | 35798 contours | polygones |

*Tableau 1 : données publiques utilisées*

Les métadonnées sont quasiment inexistantes : la date ne renseigne que la création du fichier - pas des données -, la relation entre l'objet et sa localisation n'est pas explicite : gestion des tronçons communs → écarts entre *RailwayNode*, *RailwayStationNode* et *RailwayYardNode* illustrés sur Fig.1.

Ce manque de métadonnées, dont l'impact n'a été mesuré qu'après le constat d'erreurs inexpliquées -jusqu'à 3-4 km- a conduit à l'ajout de nœuds intermédiaires afin de conserver l'information originale sans compromettre la qualité et la cohérence du réseau cible (voir section 2).

Ces sources permettent de construire le réseau ferré tel qu'aujourd'hui encore sous la responsabilité de la SNCF. Les informations sur le statut des lignes et gares (exploité, en travaux, neutralisé , en vente …, pour fret et/ou voyageurs) ne donnent malheureusement pas de dates.

De nombreuses données complémentaires – dates - et supplémentaires - lignes disparues ou non gérées par SNCF, eg : Nice-Digne, Corse -, doivent être recherchées sur d'autres sources ouvertes : (1) Wikipedia et (2) les sites volontaires de compilations de données ferroviaires (amateurs de trains). Les données manquantes sont estimées deux fois plus volumineuses, comme vu plus haut.

### 1.2.2 Wikipedia : modèles de lignes et de gares

L'encyclopédie libre Wikipedia offre des outils largement adoptés, afin de représenter des données sémantiquement structurées. En particulier l'outil *Infobox* spécialisé par catégories. Ainsi :

**Modèle:Infobox_Gare** : un extrait simplifié est fourni par le tableau 2.

```
{{Infobox Gare
 | nom             = Villaines
 | ville           = [[Villaines-sous-Bois]]
 | latitude        = 49.079542
 | longitude       = 2.350966
 | exploitant      = SNCF
 | code gare uic   = 87{{Souligner|27202}}1
 | lignes          = [[Ligne de Montsoult à Luzarches]]
 | mise en service = {{Date|1|mai|1880|dans les chemins de fer}}
 | fermeture       =
 }}
```
*Tableau 2 : modèle Gare appliqué à Villaines*

**Modèle:Infobox_Ligne_ferroviaire**: extrait simplifié en tableau 3.

```
{{Infobox Ligne ferroviaire
 | nomligne          = LGV Nord <small>(LN3)</small>
 | origine           = ([[Paris]]) [[Gonesse]]
 | destination       = [[Calais]]
 | villes            = TGV-Hte-Picardie, Lille-Europe, Calais-Fréthun
 | mise en service   = 1993
 | longueur          = 333
 | écartement        = normal
 | exploitants       = SNCF
 | schéma2           = Schéma de la LGV Nord
}}
```
*Tableau 3 : modèle Ligne-ferroviaire appliqué à LGV_Nord*

Le modèle Ligne donne accès à un schéma qui redirige vers un code particulier (Wikipedia, 2011), illustré dans l'exemple simplifié suivant (Tab.4) :

**Code 'BS-table' pour le schéma**
(BS = BahnStrecke, format d'origine germaine)

```
{{BS-table}}
{{BSbis|...|    ...  |Ligne d'Esch- ....
{{BS3bis|...|   ...  |Frontière .......
{{BS5bis|...|   ...  |Ligne de Fontoy ..
{{BS5bis|...|   ...  |(1) Ligne non ....
{{BS5bis|...| 21,144|Audun-le-Tiche | ...
{{BS5bis|...|   ...  |Ancienne voie ....
{{BSbis|...|   2,4xx|  |Bif vers .......
{{BSbis|...|   4,6xx|Rédange |(340 m)}}
{{BSbis|...|   5,5xx|  |Tunnel ... }}
{{BSbis|...|   6,6xx|  |Ancienne front..
{{BS5bis|...|   ...  |Ligne de Longwy...
{{BSbis|...|   8,8xx|Hussigny-Godbrange..
{{BSbis|..|    ...   |Ligne de Longwy...
{{BS-table-fin}}
```

*Tableau 4 : modèle BS-table appliqué à Ligne d'Esch-sur-Alzette à Audun-le-Tiche*

Si l'analyse des instances de Gare ou Ligne est simple, l'analyse des schémas est complexe : la variété des choix provoque des écarts par rapport à un usage courant, heureusement suivi à 99%.

### 1.2.3 Données ferroviaires volontaires

Pour l'accès aux lignes secondaires « oubliées » le recours réside dans la richesse des informations volontaires (Goodchild, 2010). Voici un tableau incomplet (Tab.4), avec un avantage pour la tête de liste au sujet des lignes secondaires, et au second concernant les dates des lignes SNCF fermées.

| sites | description | Fig. | espace et temps |
|---|---|---|---|
| archeoferroviaire.free.fr | Liste et tracés très exhaustifs des lignes secondaires, et autres infos | 3.a | Tracé sans coordonnées, dates |
| routes.fandom.com | Wiki très riche sur l'état des lignes principales hors service | 3.b | Pas de coordonnées, dates |
| et aussi : | | | |
| chemins.de.traverses.free.fr | | | |
| lignes-oubliees.com | | | |
| ruedupetittrain.free.fr, véritable « mémoire toponymique » | | | |
| … | | | |

*Tableau 5 : sites de données ferroviaires « volontaires »*

Contrairement à Wikipedia, ces sites ne suivent pas un modèle –format- stable et l'extraction de données nécessite des analyseurs de tableau ou de texte, adaptés (« parsing »).

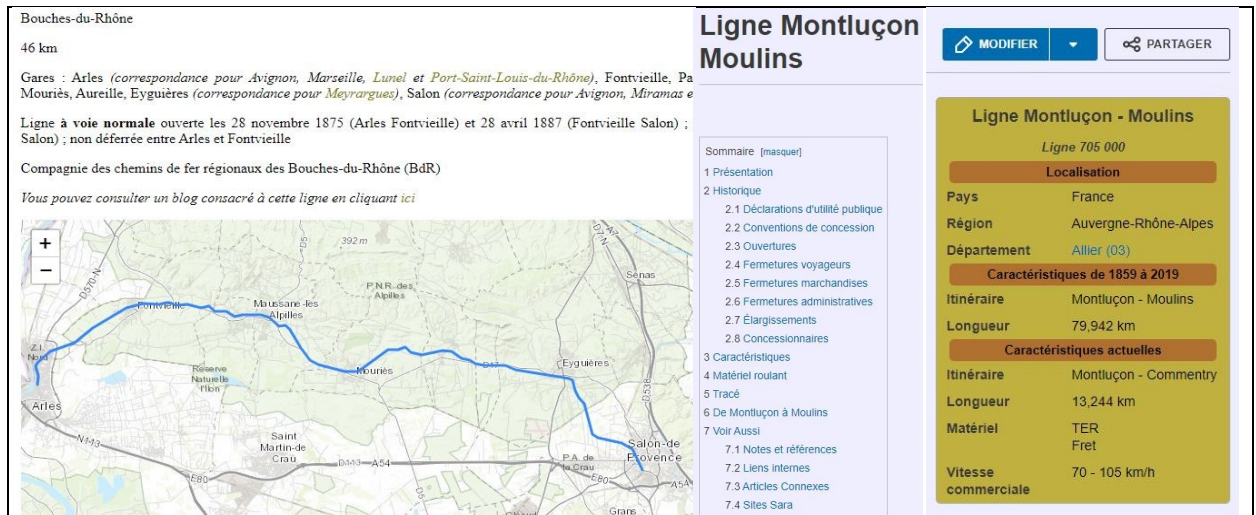

*Figure 3 :(a) exemple Arles-Salon, (b) exemple Montlucon-Moulins*

Le tracé Arles-Salon (`archeoferroviaire`) facilite la localisation des gares intermédiaires listées dans le texte. La page Montlucon-Moulins (`routes.fandom`) informe sur la date d'abandon de 83% de la ligne : 2019, pour construire un tronçon de l'autoroute RCEA Centre-Europe-Atlantique à la place.

### 1.3   Fusion de données et intégration de données

#### 1.3.1   *Fusion des données ferroviaires publiques*

La première partie du travail a consisté à « fusionner » les données issues de différentes versions SNCF, c'est-à-dire : (1) apparier les items quasi-identiques - toponymes aux accents près, ou ayant évolué dans le temps, eg : Ambérieu et Ambérieu-en-Bugey -, puis comparer les coordonnées et décider ou non, d'unifier sur la géo-localisation préférée, (2) décider de conserver les items non appariés provenant de plusieurs sources, en reconstituant les informations éventuellement absentes.

Ces opérations et leur justification théorique sont décrites dans (Bloch, 2008) et (Edwards et al., 2004). Le traitement est entièrement automatique, y compris pour isoler les cas indécis, à part.

#### 1.3.2   *Intégration de données ferroviaires volontaires*

Les gares des lignes principales ainsi fusionnées ne représentent qu'un tiers des gares ayant existé : la comparaison automatique avec les *schémas* issus de Wikipédia permet de détecter les absences et de les insérer dans le fichier cible. Ce processus, dit d'intégration, contrôle la cohérence du modèle de données et de contraintes, comme l'ordre des « points kilométrique ». Illustration en Fig 4(a) : ajout - en rouge - de gares de la « petite-ceinture » de Paris, sur résultat de fusion - en vert -.

Certaines lignes du réseau principal, carrément absentes des données SNCF, ainsi qu'un nombre significatif de lignes secondaires, sont accessibles sur Wikipedia, par transitivité des liens rencontrés, si les embranchements sont mentionnés. La même opération d'intégration est effectuée sur la ligne entière : il faut alors contrôler l'appariement des points de connexion avec une des gares existante, ou insérer un nœud -'*installation d'embranchement*', noté IE- au point de bifurcation. La seule partie manuelle consiste à localiser ce point de bifurcation sur une image ou carte numérique. Illustration Fig.4(b) de l'intégration de la ligne Pontorson-Mont-St-Michel.

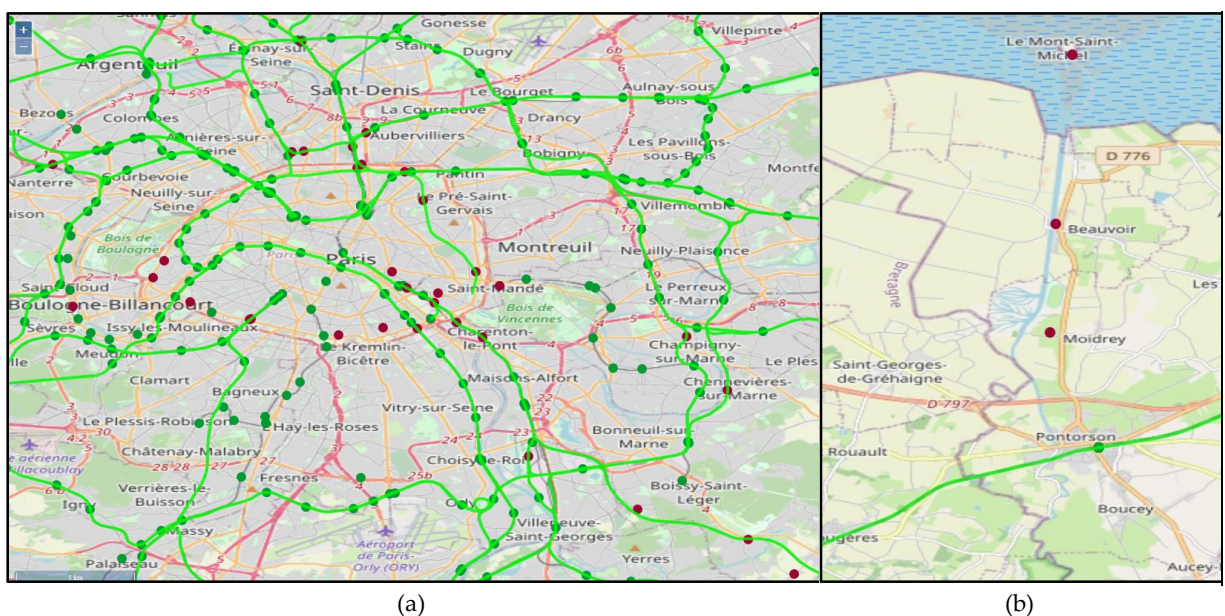

(a)                                                                 (b)

*Figure 4 : (a) ajout de gares « petite-ceinture », (b) ligne Pontorson-Mont-St-Michel.*

Les lignes secondaires et les « tramways ruraux » absents de Wikipédia, sont aussi nombreux et leur extraction se fait sur les sites volontaires (Juhàsz, 2016). La majorité de leur gares, ou simples arrêts, est listée dans le site `archeoferroviaire`, qui fournit également un tracé, de très bonne qualité, de la ligne. Certaines gares disparues sont reconnaissables sur image aérienne ancienne puis géocodées manuellement à l'aide du site : `remonterletemps.ign.fr` (IGN, 2018). D'autres cas sont plus difficiles, eg. : l'arrêt d'Irodouër du tramway d'Ille-et-Vilaine. La découverte fortuite d'une carte postale sur le site [TIV Rennes Bécherel (e-monsite.com)](#) montre l'existence d'un bâtiment qui doit être visible sur imagerie aérienne : voir figure 5.

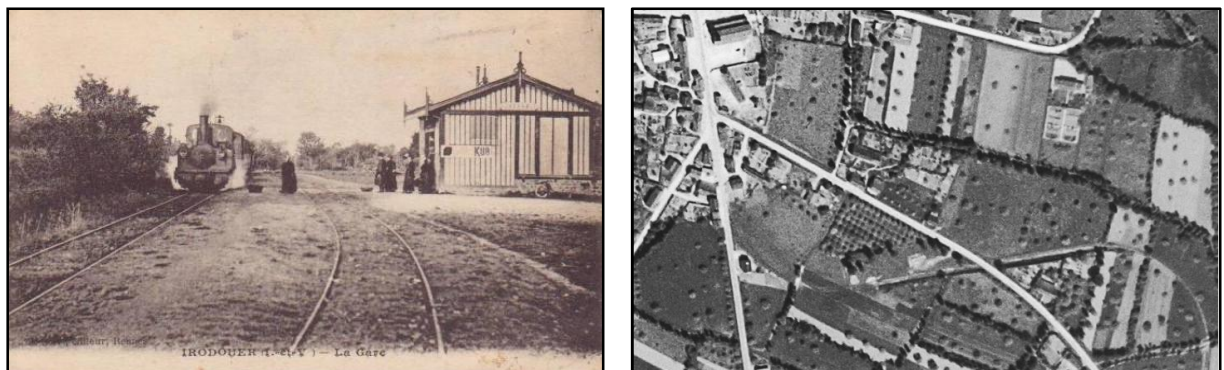

*Figure 5 : situer la ligne puis la gare d'Irodouër (test)*

Autre exemple résolu par les toponymes : la gare de Colombe-lès-Vesoul au croisement de la *rue de la gare* avec le *chemin du tacot* ! Parfois -environ 10% des gares sur lignes secondaires- le positionnement est fait manuellement à proximité d'un carrefour près de la Poste ou de la Mairie. Cette partie du travail est la plus longue pour terminer le géocodage complet. notamment pour les arrêts de tramways ruraux, difficiles à situer : l'incertitude est simplement mentionnée.

## 2  Les données reconstituées et leur qualité

### 2.1.1  Statistiques

Le jeu de données résultat est nommé CARP (Jeansoulin, 2021). A défaut de comparaison avec un échantillonnage spatial, on effectue une *méta-analyse*, terme utilisé en recherche médicale pour l'évaluation de thérapies (Riley et al., 2015). Les méta-analyses simples -à une variable- du fichier SNCF, de sa version 2014, du résultat CARP, sont comparées dans les tableaux 6 pour identifiant de gare/nœud, et 7 pour identifiant de ligne.

| Fichier | 'id' unique | 2 fois | 3 fois | 4+ fois | # nœuds | # items |
|---|---|---|---|---|---|---|
| 2014 | 5800 | 262 | 31 | 5 | 6098 | 6442 |
| 2020 | 2966 | 375 | 100 | 31 | 3472 | 4148 |
| CARP | 12986 | 1317 | 240 | 97 | 14640 | 16764 |

*Tableau 6 : méta-analyse sur l'identifiant de nœud (indique le nombre de correspondances possibles)*

| Fichier | Ligne à un nœud | % | 2- nœuds | 3-7 nœuds | 8+ nœuds | # lignes |
|---|---|---|---|---|---|---|
| 2014 | 94 | *15.8* | 62 | 191 | 246 | 593 |
| 2020 | 152 | *28.0* | 85 | 144 | 400 | 544 |
| CARP | 0 | *0* | 339 | 406 | 816 | 1561 |

*Tableau 7 : méta-analyse sur l'identifiant de ligne (indique le nombre de nœuds-gares par ligne)*

Les lignes avec nœud isolé -absurde dans la réalité- ont disparu et le nombre important de lignes avec deux nœuds s'explique par l'ajout des raccordements courts entre deux lignes (évitements utiles pour les trains de fret) : beaucoup on été fermés par économie, mais les lignes TGV en ont produit de nouveaux. A noter : on atteint quasiment le rapport 1/3 attendu.

### 2.1.2   Qualité des données

La méta-analyse multi-variée -deux paramètres ou plus- peut être réalisée sur les nœuds multiples, c'est-à-dire les gares de correspondances ou les nœuds d'embranchements, car on peut comparer les écarts entre coordonnées. La grandeur de l'écart –diamètre- est alors un indicateur de qualité. Le tableau 8 donne la répartition des diamètres des cercles de nœuds multiples($^1$) : d ≤ 150m est jugé très bon, acceptable entre 150 et 300m. Le % calculé ($^2$) concerne les 84% très bons cas : soit une très bonne qualité globale car les nombreux arrêts de tramways ruraux proches de gares de lignes principales, sont souvent au-delà des 150 m. Seuls 256 nœuds (7% des nœuds multiples) au-delà de 300m peuvent poser question : cas mentionné de *RailwayStationNode* versus *RailwayYardNode*, ou cas des gares métriques proches de gares à voie normale. Les cas graves, au-delà de 750m ont été corrigés manuellement, car peu nombreux.

| diamètre / ($^1$) | %($^2$) | [0,150m[ | [150,300m[ | [300,1km[ | [1,2km[ | [2km,∞[ |
|---|---|---|---|---|---|---|
| 2014 (*642*) | *85.7* | 550 | 48 | 32 | 6 | 6 |
| 2020 (*1182*) | *37.7* | 446 | 168 | 341 | 145 | 82 |
| CARP (*3778*) | *84.8* | 3204 | 318 | 256 | 0 | 0 |

*Tableau 8 : cohérence des coordonnées des nœuds à multiples occurences.*

Pour les gares et nœuds uniques, la qualité est estimée en fonction de celle des sources utilisées et du processus de collecte. La qualité des données Wikipedia est très bonne : la dizaine d'erreurs rencontrées concerne la séquence des gares à l'intérieur d'une ligne. La validation est faite visuellement si un tracé présente une aberration (sorte de zigzag), puis correction manuelle. La qualité des données volontaires est souvent très bonne (amateurs passionnés), mais le géocodage précis des gares n'est pas toujours possible sans trace visible, ou sans autre information.

### 2.1.3   affichage des résultats bruts

Le fichier au format « geojson », avec coordonnées [longitude,latitude], peut être lu par tout afficheur dédié (eg. *geojsonlint.com* sur Fig.6-gauche), mais il vaut mieux définir des « styles » adéquats : Fig.6-droite, avec OpenLayers. Si le tracé d'une ligne existe dans les données SNCF, il est utilisé. Sinon on affiche des segments entre nœuds successifs (cf. Fig.7).

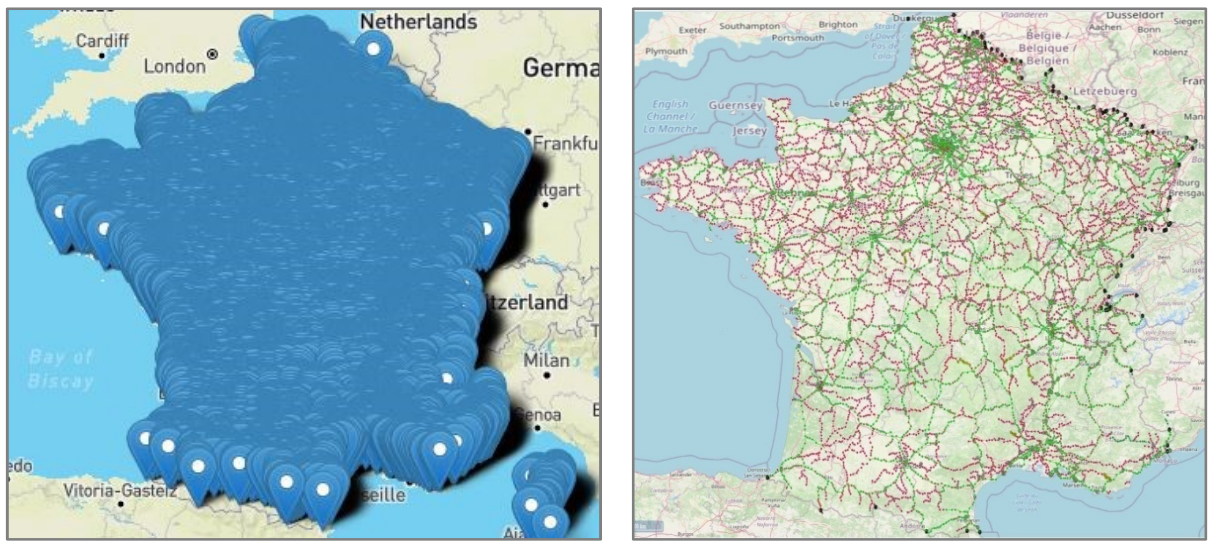

*Figure 6 : affichage par marqueurs standard, versus style adapté à la résolution.*

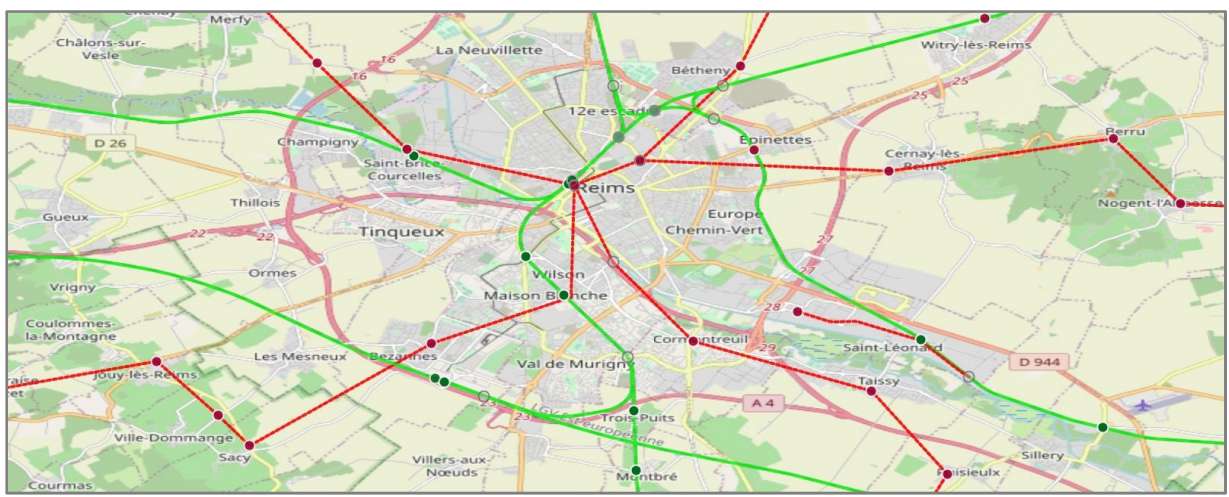

*Figure 7 : affichage des gares (rouge ou vert si en service), des nœuds (cercle gris), et des lignes avec tracé réel ou simulé par segment. Logiciel libre OpenLayers. Fond de carte OpenStreetMap.*

## 3   Evaluation et simulation : outils et perspectives

### 3.1   Les communes servies par le réseau ferroviaire

A ce jour -juillet 2021- les données sont complètes sur la moitié Nord de la France, et autour de 80% sur la moitié Sud. Seules les voies ferrées d'écartement standard ou métrique sont enregistrées et les voies fermées avant 1925 sont ignorées. Les communes françaises ont peu évolué durant le siècle dernier, à l'exception d'une politique de « fusion » depuis une dizaine d'années. Le fichier des polygones de communes donne approximativement la situation de 2015.

La figure 8 affiche les communes qui possèdent une gare recensée par le fichier SNCF-2020 (haut) et celles correspondant au cumul centennal du réseau reconstitué (bas), sur une même zone Nord-Ouest de la France.

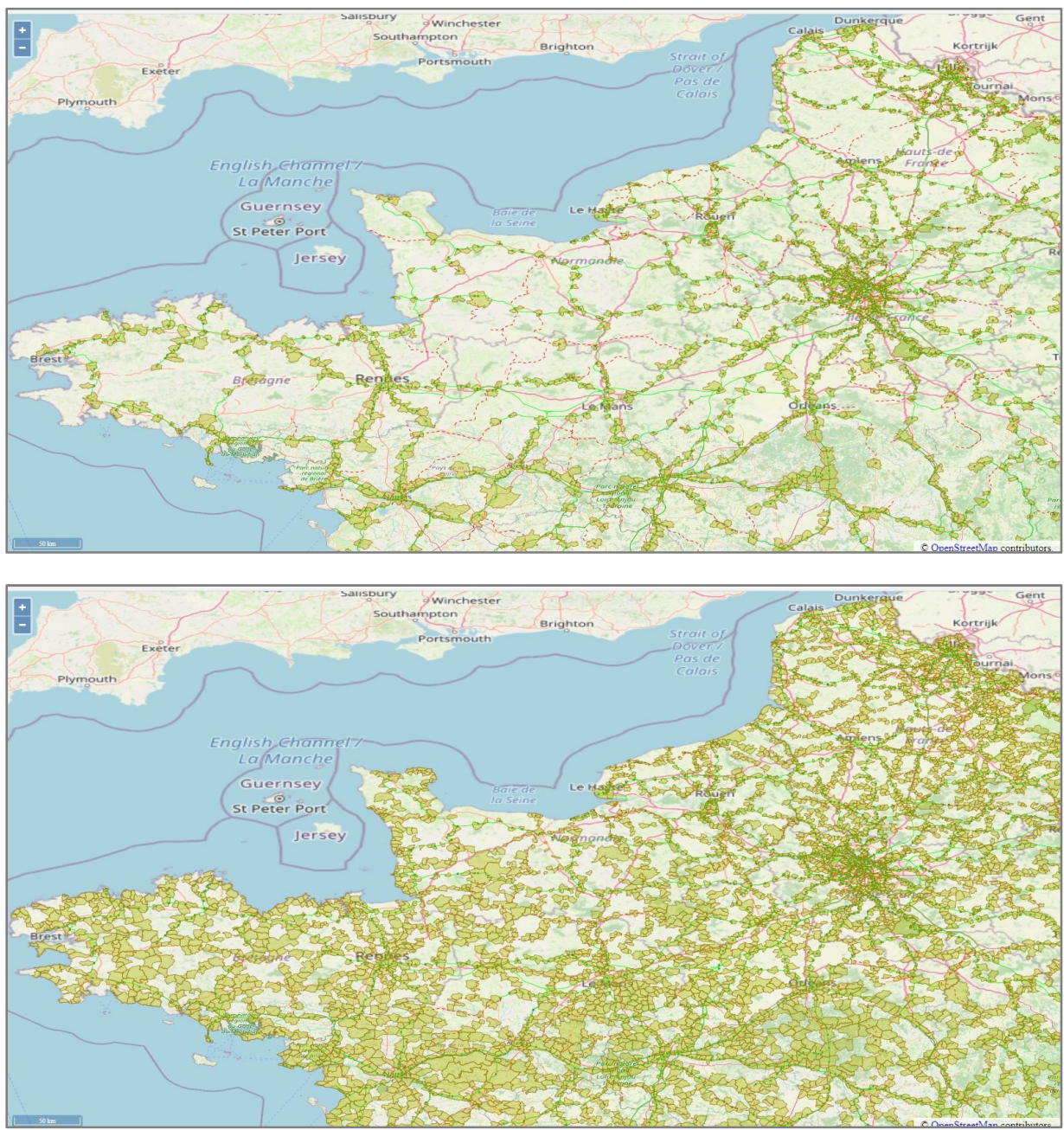

*Figure 8 :affichage des communes desservies en 2020 (haut), ou au cours du siècle écoulé (bas).*

**Résultat** : les gares recensées par SNCF-2020 correspondent à 3112 communes (8,7 %), les gares du réseau reconstitué – à ce jour – correspondent à 10237 communes (28,6 %) : rapport 1:3 attendu.

**Commentaires** : le maillage territorial est dense, à quelques exceptions. Le département du Calvados et les plages normandes semblent mal desservies : c'est dû au choix de ce département de financer des lignes à écartement de 60 cm, non répertoriées dans cette reconstitution. Autre département déficitaire, l'Aube, où les travaux ont démarré en 1912 et n'ont jamais repris après la fin de la guerre.

Perspectives possibles : la liste des communes donne accès aux superficies de territoires, aux groupements par terroir ou tout autre « bassin » d'activité pertinent. Elle donne accès à la démographie et à l'activité économique au cours de la période considérée : nombreuses applications potentielles en perspective.

### 3.2 Calcul de trajets, de coûts de transport et de coût environnemental

Un graphe, au sens mathématique, donne accès à de nombreux algorithmes, notamment au calcul de tous les chemins possible, dont le meilleur chemin ou des chemins alternatifs si on ajoute une fonction de coût (eg. : le plus rapide selon vitesse autorisée, le moins coûteux selon dénivelé total, etc.). Pour passer du fichier geojson actuel à un graphe mathématique, un transcodage sera réalisé prochainement, mais on peut déjà anticiper son utilisation.

#### *3.2.1 Fonction de coût énergétique, de coût d'émission de $CO^2$*

Le coût énergétique d'un moyen de transport revient au coût de la traction qui permet de vaincre les forces de frottement. Le diagramme GvK (Fig.9) compare l'efficacité des forces de traction de divers véhicules -terrestres, aériens ou maritimes- en fonction de la vitesse. Ce diagramme, originalement de 1950 (Gabrielli & von Kármán, 1950), a été mis à jour par (Yong et al., 2004), qui montre que le train est le moyen de transport terrestre le moins énergivore et celui qui a fait le plus de progrès énergétiques.

La raison est bien connue : les faibles frottements sont dûs au très faible coefficient d'adhérence roue-rail (Alacoque et al., 2222). Ceci a des inconvénients (distance de freinage, risque de patinage), mais des qualités exceptionnelles : le coût de la tonne au kilomètre est d'autant plus intéressant que le poids est plus important et que le train est plus long.

L'article L. 1431-3 du code des transports impose aux entreprises du secteur d'informer sur les émissions de gaz à effet de serre de leur prestation. Le Ministère des Transports a établi, en accord avec les professionnels, un Guide méthodologique (MET, 2012), pour effectuer un calcul de coût $CO^2$ pour tout couple (origine, destination), selon les caractéristiques du chemin utilisé.

Le tableau suivant (Tab.9) présente quelques données extraites de ce guide méthodologique :

| Moyen de transport | Energie | Charge moyenne tenant compte des trajets à vide | Taux de consommation |
|---|---|---|---|
| Rail : train | Electricité | de 400-600 tonnes | 16,7 kWh / km |
| | Gazole non routier | | 3,87 kg / km |
| Camion 19 t. | Gazole routier | 2,5 tonnes | 0,27 l / km |
| Camion 40 t. | Gazole routier | 6 tonnes | 0,34 l / km (environ 29 kg / km) |

*Tableau 9 : consommation énergétique par tonne transportée*

Transporter 600 tonnes par camions consomme 29 kg / km, soit 7,4 fois plus que par train gazole !

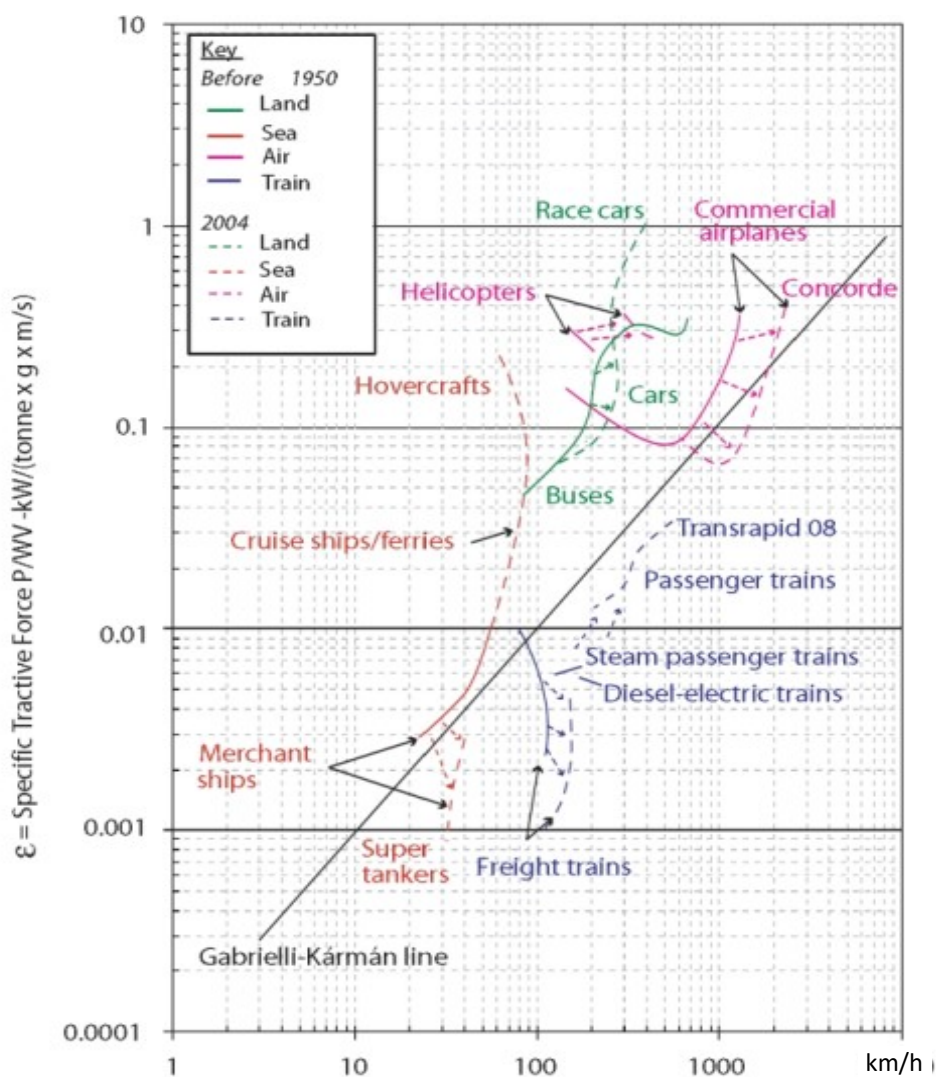

*Figure 9 : diagramme GvK « revisité », par Imperial College's Railway Research Group.*

### *3.2.2 Simulation de coût environnemental*

Le guide méthodologique fournit les taux d'émission de $CO_2$ par type de transport (Tab.10).

| Moyen de transport | Energie | Utilisation plus ou moins énergivore | Taux d'émission $CO_2$ |
|---|---|---|---|
| Rail : train | Electricité (France) | Fret pondéreux à léger | 1,47 à 2,2 g $CO_2$ / t.km |
|  | Gazole non routier |  | 23 à 35 g $CO_2$ / t.km |
| Camion 19 t. | Gazole routier | pas de variation notable | 332 g $CO_2$ / t.km |
| Camion 40 t. | Gazole routier | citerne – frigorifique | 83 à 105 g $CO_2$ / t.km |

*Tableau 10 : taux d'émission (inclut phase amont + consommation de l'énergie)*

L'électricité produite en France, associée au transport des marchandises lourdes par train, est, de loin, la meilleure solution pour réduire les émissions à effet de serre produites par le transport de marchandise : de 40 à 50 fois moins polluant ! Même le transport léger de produits réfrigérés sur une ligne non électrifiée est 3 fois moins polluant par train, et 30 fois moins sur ligne électrifiée.

Les simulations possibles sont innombrables avec les outils ainsi disponibles, par exemple pour le Perpignan-Rungis dont la suppression est en débat, ou pour évaluer le coût environnemental des dés-électrifications de certaines lignes pyrénéennes.

## Conclusion

Seul le « temps long » permet de comprendre l'effet global de décisions politiques échelonnées sur plusieurs générations. Dans les années 50 nous n'avions pas les connaissances sur les gaz à effet de serre, mais depuis les années 70 oui, et l'inertie politique est bien plus lourde à déplacer qu'un train de 600 tonnes ? 30% des communes françaises ont eu accès au chemin de fer dans le siècle dernier ; contre 8% aujourd'hui, concentrées autour des grandes métropoles et quelques grands axes. Voilà une des premières conclusions quantifiée du jeu de données décrit dans cet article.

Ce jeu de données permet déjà non seulement de regarder en arrière, mais aussi d'appréhender la tendance des décisions probables et d'anticiper si des changements sont possibles. En tous cas permet de mesurer objectivement le coût de certaines décisions en débat, concernant les fermetures ou réouvertures de lignes ferroviaires.

A ce jour la collecte des gares et leur géocodage sont terminés à 90% environ. Concernant la datation des mises hors service des lignes, il faudrait l'étendre au niveau des tronçons de ligne pour mieux comprendre l'effet de fermetures successives partielles qui deviennent irréversibles : l'exemple de la route transversale « Centre Europe-Atlantique » RCEA, mériterait d'être illustré par l'étude d'une telle succession.

Ce travail n'est pas seulement franco-français. Les outils de reconstitution présentés en section 1 ont été appliqués avec succès pour des lignes frontalières en Belgique et Luxembourg et pourraient être adaptés aux variantes de formats rencontrés pour d'autres pays européens. Ceci permettrait certaines études comparatives et une meilleure compréhension des dynamiques frontalières.

### Jeu de données : accès en ligne

Le jeu de données peut être visualisé en ligne et de manière interactive sur le site : http://bigbugdata.com/cartorail.html (fond de carte OpenStreetMap). Les diverses couches peuvent être sélectionnées individuellement, gares et lignes sont 'clicables'. Travail en cours d'amélioration.

### Références